\documentclass[preprint,showpacs,preprintnumbers,amsmath,amssymb]{revtex4}

\usepackage{graphicx}
\usepackage{dcolumn}
\usepackage{bm}
\usepackage{epsf}

\renewcommand{\bar}[1]{\overline{#1}}

\renewcommand{\d}{{\mathrm d}}

\usepackage{amssymb}

\renewcommand{\d}{{\mathrm d}}
\renewcommand{\bar}[1]{\overline{#1}}

\usepackage{indentfirst}

\providecommand{\Journal}[4] {#1 {\bf #2}, #3 (#4)}
\providecommand{\NPA}{Nucl. Phys. A } %
\providecommand{\NPB}{Nucl. Phys. B } %
\providecommand{\PRL}{Phys. Rev. Lett. } %
\providecommand{\PRC}{Phys. Rev. C } %
\providecommand{\PRD}{Phys. Rev. D } %

\newcommand{\be}{\begin{equation}}
\newcommand{\ee}{\end{equation}}
\newcommand{\ba}{\begin{eqnarray}}
\newcommand{\ea}{\end{eqnarray}}

\newcommand{\ra}{\rangle}

\newcommand{\sigmaPiN}{\sigma_{\pi N}}
\def\ra{\rightarrow}

\def\be{\begin{equation}}
\def\ee{\end{equation}}
\def\bea{\begin{eqnarray}}
\def\eea{\end{eqnarray}}

\begin{document}

\title{Phenomenological analysis of nucleon strangeness and meson-nucleon sigma terms}

\author{Feng Huang}
\affiliation{Department of Physics, Peking University, Beijing
100871, China} \affiliation{National Astronomical Observatories,
Chinese Academy of Sciences, Beijing 100012, China}
\author{Fu-Guang Cao}
\affiliation{Institute of Fundamental Sciences, Massey University, Private Bag 11 222, Palmerston North, New Zealand}
\author{Bo-Qiang Ma}
\email{mabq@phy.pku.edu.cn} 
\affiliation{Department of Physics, Peking University, Beijing
100871, China\footnote{Mailing address.}} \affiliation{State Key
Laboratory of Nuclear Physics and Technology, Peking University,
Beijing 100871, China}

\begin{abstract}
We calculate the nucleon strangeness $y_N$ in the chiral quark model
and the meson cloud model. With the internal relation between the
sigma term of $\pi$$N$~($\sigma_{\pi N}$) and $y_N$, we present the
results of $\sigma_{\pi N}$ in these two models. Our calculations
show that $y_N$ from the chiral quark model is significant larger
than that from the meson cloud model, whereas the difference of
$\sigma_{\pi N}$ between the two models is relatively small. We also
present the results of $\sigma_{K N}$ and $\sigma_{\eta N}$, which
could be determined by $\sigma_{\pi N}$ and $y_N$ from their
definition in the current algebra, and find that these two physical
parameters are quite sensitive to $y_N$. The results indicate the
necessity to restrict the parameters of the two models from more
precision measurements.

\end{abstract}

\pacs{13.75.Gx, 12.39.Fe, 13.75.Jz, 14.40.Aq} \vfill \maketitle

\par

\section{Introduction }
The nucleon structure has received much attention for its abundant
phenomena away from naive theoretical expectations. One of them is
that the nucleon may contain a significant component of
strange-antistrange ($s\bar{s}$) pairs. Recently, a
strange-antistrange asymmetry~\cite{ST,bm96} has been applied to
explain the NuTeV anomaly~\cite{zell02}, while the asymmetric
strange-antistrange contribution from the chiral quark
model~\cite{dm04,ding,wakamastu04} is predicted to be different from
that in the meson cloud model~\cite{Cao}. Apart from possible
high-energy experimental verifications, some parameters measured in
low-energy hadron physics could provide an instructive restriction
on the nucleon strangeness which is usually indicated by the ratio
of the strange component over the light components, $y_N$. The first
example of a large strange component in the nucleon was from the the
measurement of the pion-nucleon sigma term~($\sigma_{\pi N}$)
\cite{TPCheng1976} which is a fundamental parameter of low-energy
hadron physics. The precise value of $\sigmaPiN$ is of practical
importance for numerous phenomenological applications. It provides a
direct test of chiral symmetry breaking effects and effective quark
models since this quantity is sensitive to the quark-antiquark sea
contribution \cite{review-early}.

The $\sigma_{\pi N}$ can not be measured directly
\cite{review-early,review-recent}. There are two ways discussed in
the literatures to determine the value of $\sigma_{\pi N}$
experimentally. The combination $\hat{\sigma}=\sigmaPiN(1-2y_N)$
which measures the strength of the matrix element $\langle p |
\bar{u}u+\bar{d}d-2\bar{s}s|p \rangle$ can be related to the baryon
mass differences in the SU(3) limit \cite{TPCheng1976}. Thus the
$\sigma_{\pi N}$ can be ``measured" from an analysis of the baryon
mass spectrum using chiral perturbation theory (ChPT) and
information on the value of $y_N$. A value about $45$ MeV was
obtained in Refs.\cite{g1,bm}. The other method is to  relate the
value of the scalar-isoscalar form factor $\sigma(t)$ at the point
$t=2m_\pi^2$ to the isospin-even pion-nucleon scattering amplitude
by using dispersion relation and phase shifts
\cite{Koch:pu,Gasser:1990ce}. Earlier analyses of the pion-nucleon
scattering data by Koch \cite{Koch:pu} and Gasser et al.
\cite{Gasser:1990ce} gave for $\sigma(2m_\pi^2)$ a value about
60~MeV. Using $\sigma(2m_\pi^2)-\sigma(0)=15 \,{\rm MeV}$ found by
Gasser et al.\ \cite{Gasser:1990ce}, one obtains for
$\sigmaPiN\equiv\sigma(0)$ a value around $45\,{\rm MeV}$ which
agrees with the value obtained from the baryon mass spectrum.
Several updated analyses \cite{Kaufmann:dd,Olsson:pi}, however,
tended to yield higher values for $\sigma(2m_\pi^2)$ in the range of
$(70 \sim 90)\,{\rm MeV}$,
which resulted in a value of $\sigmaPiN$ much larger than earlier analyses.
Meanwhile, recent lattice calculations gave a value in the range of $33 \sim
50\,{\rm MeV}$ depending on the extrapolation ansatz~\cite{Leinweber:2003dg}
and $43 \sim 49 \, {\rm  MeV}$ \cite{MProcuraHW}.

In this paper, we present a theoretical calculation of the nucleon strangeness $y_N$
using the chiral quark model~(CQM) and the meson cloud model~(MCM) which both have been applied to
the study of the nucleon structure extensively.
These models provide different mechanism to incorporate the meson degree of freedom in the nucleon
and the contributions from different mesons can be easily recognized.
In order to get a relatively accurate estimation on the value of $y_N$, we
calculate the contributions to the non-perturbative sea from various mesons.
Then, we use $y_N$ as an input to calculate the result of $\sigma_{\pi
N}$. Moreover, we present the results of $\sigma_{K N}$ and
$\sigma_{\eta N}$ which can be easily derived from
$\sigma_{\pi N}$ and $y_N$.

The paper is organized as follows. The definition of
the sigma terms and the internal relation between the nucleon sigma terms
($\sigma_{\pi N}$, $\sigma_{K N}$ and $\sigma_{\eta N}$) and $y_N$ are given in Section II.
The basic formalism in the CQM and MCM used to obtain
the non-perturbative strange quark component is presented in Sections III and IV, respectively.
Numerical results are given in Section V and the last section is reserved for a summary.

\section{The sigma terms and The relation between $y_N$ and $\sigma_{\pi N}$}

The pion-nucleon sigma term is defined as \cite{review-early,review-recent}
\begin{eqnarray}
\sigma_{\pi N} = \hat{m} \langle N |
\bar{u}u+\bar{d}d| N \rangle\,, \hspace{1.cm}
\label{definition}
\end{eqnarray}
where terms proportional to
$(m_u - m_d)  \langle N | \bar{u}u+\bar{d}d-2\bar{s}s| N \rangle $ have been neglected, and
$\hat{m}=\frac{1}{2}(m_u+m_d)$ is the average value of current quark masses of the $u$ and $d$ quarks.
The scalar operator $\bar{q} q$ measures the sum of the quark and antiquark numbers \cite{TPCheng1976},
thus $\sigma_{\pi N}$ is the contribution to the nucleon mass from the $u$ and $d$ quarks having a mass of $\hat{m}$.
Algebraically, $\sigma_{\pi N}$ can be written in the form of
\begin{eqnarray}
\sigma_{\pi N} = \hat{m} \frac{\langle p |
\bar{u}u+\bar{d}d-2\bar{s}s|p
\rangle}{1-2y_N}=\frac{\hat{\sigma}}{1-2y_N} \,,
\end{eqnarray}
where
\begin{eqnarray}
\hat{\sigma}= \hat{m} \langle p |
\bar{u}u+\bar{d}d-2\bar{s}s|p \rangle,
\end{eqnarray}
and
\begin{eqnarray}
y_N = \frac{\langle p | \bar{s}s |p \rangle}{\langle p | \bar{u}u
+ \bar{d}d |p \rangle}
\end{eqnarray}
represents the strangeness content of the nucleon.

Moreover, $\sigma_{K N}$ and $\sigma_{\eta N}$ can be defined in
the same way as \cite{SigmaTerm01}
\begin{eqnarray}\label{sigma_all}
& &\sigma_{KN}^{u} \equiv
\frac{\hat m + m_s}{2} \langle p|\bar u u + \bar s s|p\rangle\,, \nonumber\\
& &\sigma_{KN}^{d} \equiv
\frac{\hat m + m_s}{2} \langle p|\bar d d + \bar s s|p\rangle\,, \nonumber\\
&
&\sigma_{KN}^{I=0}\equiv\frac{\sigma_{KN}^u+\sigma_{KN}^d}{2}\,, \nonumber\\
& &\sigma_{\eta N} \equiv \frac{1}{3} \langle p|\hat m (\bar u
u + \bar d d)  + 2 m_s \bar s s  |p\rangle\,.
\end{eqnarray}
The $\sigma_{K N}$ and $\sigma_{\eta N}$ can be expressed in term of $\sigma_{\pi N}$ and $y_N$,
\begin{eqnarray}
& &\sigma_{KN}^{I=0}=\sigma_{\pi N}(1+2y_N)
\frac{\hat m+m_s}{4\hat m}=\frac{13}{2}\sigma_{\pi N}(1+2y_N)\,,  \\
& &\sigma_{\eta N}=\sigma_{\pi N} \frac{\hat m + 2 y_N m_s}{3\hat
m}=\sigma_{\pi N} \frac{1 + 50y_N}{3}.
\end{eqnarray}
where the quark mass ratio $m_s / \hat{m}\simeq 25$ \cite{ratio25} has been used.

The ``quark flavor fraction'' in a nucleon, $f_q$, was defined by
Cheng and Li \cite{cl1,cl2},
\begin{eqnarray}
f_q = \frac{ \langle p | \bar{q}q |p \rangle}{\langle p | \bar{u}u
+ \bar{d}d+ \bar{s}s |p \rangle}=\frac{ q+\bar{q}}{3+2( \bar{u} +
\bar{d}+ \bar{s})}\,,
\label{EQ_fq}
\end{eqnarray}
where $q$ and $\bar{q}$ in the proton matrix elements $\langle p | \bar{q}q |p
\rangle$ are the quark and antiquark field operators, and in the last term they
stand for the quark and antiquark numbers in the nucleon.
Thus, one can express $y_N$ in term of quark and antiquark numbers in the nucleon
and the strange quark fraction $f_s$,
\begin{eqnarray}
y_N=\frac{2\bar{s}}{3+2(\bar{u} + \bar{d})} \,
\end{eqnarray}
and
\begin{eqnarray}
y_N=\frac{f_s}{1-f_s}.
\label{EQ_yNfs}
\end{eqnarray}
It is worth pointing out that the quark and antiquark numbers in Eqs.~(\ref{EQ_fq}) and (\ref{EQ_yNfs})
are the total numbers in the nucleon.

As discussed in the Introduction, we need to calculate $y_N$ directly
in the two models, which will avoid discussing the much more
complicated case about the matrix elements with mass parameters and reduce
the model dependence of calculation since the quantity calculated is a ratio.
For the $\hat{\sigma}$ part, it was normally adopted as
$\hat{\sigma} \simeq 26  \; {\rm MeV} $ in leading order in the ChPT  \cite{review-early}, while
a larger value $ \hat{\sigma} = 33 \pm 5 \; {\rm MeV}$
was obtained by Gasser and Leutwyler \cite{g1} in $O(m_q^{3/2})$ calculation.
Borasoy and Mei\ss ner \cite{bm} analyzed the octet baryon masses in the heavy baryon framework
of ChPT to order $O(m_q^2)$ and obtained $ \hat{\sigma} = 36 \pm 7 \; {\rm MeV}$.
In this paper, we will perform calculation with $\hat{\sigma}=26$ MeV and $36$ MeV respectively,
and compare the effects on the sigma terms.

\section{Chiral quark model}
The chiral quark model which was first formulated by Manohar and Georgi
\cite{mg} describes successfully the nucleon properties in the
scale range from $\Lambda_{\rm QCD}$ ($0.2 \sim 0.3$ GeV) to
$\Lambda_{\chi{\rm SB}}$ ($\sim$ 1 GeV). The dominant interaction is
the coupling among constituent (dressed) quarks and Goldstone
bosons (GBs), while the gluon effect is expected to be small. This model
has been employed in the study of flavor asymmetry in the nucleon sea \cite{ehq92} and
the proton spin problem by introducing SU(3) breaking and U(1)
breaking effects \cite{cl1,cl2,AJBuchmannF96,smw,wsk,song9705}.

The effective Lagrangian describing interaction between quarks and
the nonet of GBs can be expressed as \cite{cl1}
\begin{eqnarray}
L_I= g_8 \bar{q}\Phi q + g_1\bar{q}\frac{\eta'}{\sqrt 3}q = g_8
\bar{q}\left( \Phi+ \zeta \frac{\eta'}{\sqrt 3}I \right) q \,,
\end{eqnarray}
where $\zeta=g_1/g_8$, $g_1$ and $g_8$ are coupling constants
for the singlet and octet GBs, respectively, and $I$ is the $3\times
3$ identity matrix.  The GB field which includes the octet and the
singlet GBs is written as
\begin{eqnarray}
L_I=g_8 \bar{q} \left(
\begin{array}{ccc}
\frac{\pi^0}{\sqrt 2}
+\beta\frac{\eta}{\sqrt{6}}+\zeta\frac{\eta'}{\sqrt{3}}& \pi^+ &
\alpha K^+ \\ \pi^- & -\frac{\pi^0}{\sqrt 2}
+\beta\frac{\eta}{\sqrt{6}}+\zeta\frac{\eta'}{\sqrt{3}} & \alpha
K^0 \\ \alpha K^- & \alpha \bar{K}^0 & -\beta
\frac{\eta}{\sqrt{6}}+\zeta \frac{\eta'}{\sqrt{3}}\\
\end{array}
\right )q
\,.
\label{EQ_Leff}
\end{eqnarray}
The SU(3) symmetry breaking is introduced by considering
$m_s >m_{u,d}$ and the masses of GBs to be non-degenerate $(M_{K,\eta} > M_{\pi})$,
whereas the axial U(1) breaking is introduced by $M_{\eta^{'}} > M_{K,\eta}$.
These effects are expressed in Eq.~(\ref{EQ_Leff}) by suppression factors $\beta$ (for $\eta$),
$\alpha$ (for $K$), and $\zeta$ (for $\eta'$) which deviate
from $1$, the value for the symmetric limit.
These values are generally fixed by the experimental data of the
light antiquark asymmetry and quark spin component.
The probabilities of chiral fluctuations $u(d) \rightarrow d(u) + \pi^+(\pi^-)$,
$u(d) \rightarrow s + K^{+} (K^0)$, $u(d,s) \rightarrow u(d,s) + \eta$, and $u(d,s) \rightarrow u(d,s) + \eta^{'}$
are proportional to $a(=|g_8|^2)$, $\alpha^2 a$, $\beta^2 a$ and $\zeta^2 a$, respectively.
The antiquark numbers calculated using the effective Lagrangian Eq.~(\ref{EQ_Leff}) are \cite{cl2}
\begin{eqnarray}
\bar u &=&\frac{1}{12}[(2 \zeta+\beta+1)^2 +20] a\,, \nonumber\\
\bar d &=&\frac{1}{12}[(2 \zeta+ \beta -1)^2 +32] a\,, \nonumber \\
\bar s &=&\frac{1}{3}[(\zeta -\beta)^2 +9 {\alpha}^{2}] a\,.
\label{eq:qbar1}
\end{eqnarray}

Another method to introduce the symmetry breaking effects is to calculate directly the
different fluctuations. The antiquark distribution functions are given by~\cite{sw98}
\begin{equation}
\bar{q}_{k}(x)=\int\frac{\textmd{d}y_{1}}{y_{1}}\frac{\mathrm d
y_{2}}{y_{2}}V_{\bar{k}/\delta}(\frac{x}{y_{1}})P_{\delta j/i}(\frac{y_{1}}{y_{2}})q_{i}(y_{2})\,,
\label{EQ_qbarCQM2}
\end{equation}
where $V_{\bar{k}/\delta}(x)$ is the antiquark $\bar{k}$ distribution function
in a Gladstone Boson $\delta$.
$P_{\delta j/i}(y)$ is the splitting
function giving the probability for a parent quark $i$ to split into a quark $j$ with the light-cone momentum fraction $y$
and transverse momentum $\emph{\textbf{k}}_\bot$,
and a spectator GB~($\delta=\pi, K, \eta$) with the light-cone momentum fraction $1-y$ and transverse momentum
$\emph{\textbf{k}}_\bot$,
\begin{eqnarray}
P_{j \delta /i}(y)=\frac{1}{8\pi^2}(\frac{g_{A}\bar{m}}{f})^2\int
\textmd{d}k^{2}_{\bot}\frac{(m_j-m_{i}y)^2+k^{2}_{\bot}}{y^{2}(1-y)
[m_{i}^{2}-M^{2}_{j\delta}]^{2}} \,,
\label{EQ_PCQM2}
\end{eqnarray}
and
\begin{eqnarray}
P_{\delta j/i}(y)=P_{j \delta /i}(1-y).
\end{eqnarray}
In Eq.~(\ref{EQ_PCQM2}), $m_{i}$ and $m_{j}$ are  the masses of the $i,
j$-constituent quarks, $\bar{m}=(m_{i}+m_{j})/2$ is the average mass of the constituent quarks,
and $m_{\delta}$ is the mass of the GB.
$M^{2}_{j\delta}$ is the invariant mass squared of the final state,
\begin{eqnarray}
M^{2}_{j\delta}=\frac{m^{2}_{j}+k^{2}_{\bot}}{y}+\frac{m^{2}_{\delta}+k^{2}_{\bot}}{1-y}.
\end{eqnarray}

The antiquark numbers can be obtained by integrating $\bar{q}(x)$ given by Eq.~(\ref{EQ_qbarCQM2}) over $x$,
\begin{eqnarray}
\bar{u}&=&\frac{7}{4}\langle P_{\pi^+}\rangle +
\frac{1}{12}\langle P_{\eta}\rangle + \frac{1}{3}\langle P_{\eta'}\rangle \,,\nonumber \\
\bar{d}&=&\frac{11}{4}\langle P_{\pi^+}\rangle +
\frac{1}{12}\langle P_{\eta}\rangle +\frac{1}{3}\langle P_{\eta'}\rangle \,,\nonumber \\
\bar{s}&=&3\langle P_{k^+}\rangle + \frac{4}{3}\langle
P_{\eta}\rangle +\frac{1}{3}\langle P_{\eta'}\rangle \ ,
\label{eq:qbar2}
\end{eqnarray}
where
$\langle P_{\delta}\rangle \equiv \langle P_{j\delta/i}\rangle=\langle P_{\delta j/i}\rangle$
is the first moment of splitting functions~\cite{sw98}.
The contributions from $\eta'$ meson are included in Eq.~(\ref{eq:qbar2}) in order to compare with
results from Eq.~(\ref{eq:qbar1}).

An exponential form factor is usually introduced in the calculation,
\begin{equation}
g_{A}=g_{A}^{\prime}\textmd{exp}\bigg{[}\frac{m^{2}_{i}-M^{2}_{j\delta}}{4\Lambda_{\delta}^{2}}\bigg{]},
\end{equation}
with $g_{A}^{\prime}=1$ following the large $N_{c}$ argument
\cite{w90}.
The cutoff parameter $\Lambda_{\delta}$ which can be
determined by fitting the Gottfried sum rule was taken to be
$\Lambda_{\delta}=\Lambda_{\pi}=1500$~MeV in \cite{sw98}.
Recently, this method was used to calculate the
strange-antistrange asymmetry \cite{ding} in which
$\Lambda_{K}=(900 \sim 1100)$~MeV is adopted.
The dependence of numerical calculations on $\Lambda_{\delta}$ is studied by
taking $\Lambda_{\delta}=1100$~MeV and $1500$~MeV.
The mass parameters are taken to be
$m_{u}=m_{d}=330$~MeV, $m_{s}=480$~MeV,
$m_{\pi^{\pm}}=m_{\pi^{0}}=140$~MeV,
$m_{K^{+}}=m_{K^{0}}=495$~MeV, $m_{\eta}=548$~MeV and
$m_{\eta'}=958$~MeV.

\section{Meson Cloud Model}

The meson cloud model is very successful in explaining many non-perturbative
properties of the nucleon \cite{first,first2,MA,Review,KoepfFS,HolzenkampHS+Holtmann}.
In the meson cloud model, the nucleon can be viewed as a
bare nucleon~(core) plus a series of baryon-meson Fock states
which result from the nucleon fluctuating into a baryon plus a meson
$N\rightarrow BM$~(a bare core surrounded by a meson cloud).
The physical nucleon wave function is composed of various baryon-meson Fock states
\begin{eqnarray}
\left|N\right>=\sqrt{Z}\left|N\right>_{\tiny{\textrm{bare}}}+\sum_{BM}\int
\d y \d ^{2}k_{\perp}\Psi_{BM}(y,k_{\bot}^{2})
\left|B(y,\emph{\textbf{k}}_{\bot}),M(1-y,-\emph{\textbf{k}}_{\bot})\right> \, ,
\end{eqnarray}
where $Z$ is the wave function renormalization constant, and
$\Psi_{BM}(y,k_{\bot}^{2})$ is the probability amplitude for
finding a physical nucleon in a state consisting of a baryon $B$
with longitudinal moment fraction $y$ and transverse momentum
$\emph{\textbf{k}}_{\bot}$, and a meson $M$ with longitudinal
moment fraction $1-y$ and transverse momentum
$-\emph{\textbf{k}}_{\bot}$.

The model assumes that the life-time of a virtual baryon-meson Fock
state is much longer than the interaction time in deep inelastic
scattering~(DIS) or Drell-Yan processes, thus the quark and
anti-quark in the virtual baryon-meson Fock states can contribute
to the parton distribution of the nucleon.
The quark distribution $q(x)$ in the nucleon is given by
\begin{eqnarray}
q(x)=Z q_{\tiny{\textrm{bare}}}(x)+\delta q(x),
\label{Eq:qtotal_MCM}
\end{eqnarray}
where $q_{\tiny{\textrm{bare}}}$ and $\delta q$ are the contributions from the bare nucleon and
the meson and baryon cloud.

The contribution $\delta q(x)$ can be calculated via a convolution between the
fluctuation function which describes the microscopic process
$N\rightarrow BM$, and the quark~(anti-quark) distribution of
hadrons in the Fock states $\left|BM\right> $,
\begin{eqnarray}
\delta q(x)=\sum_{MB}\left[ \int_x^1\frac{\d
y}{y}f_{MB}(y)q_{M}(\frac{x}{y})+\int_x^1\frac{\d
y}{y}f_{BM}(y)q_{B}(\frac{x}{y}) \right]
\label{eq:qbarMCM}
\end{eqnarray}
where $q_M$ and $q_B$ are the quark distributions in the cloud meson and baryon respectively and
$f_{BM}$ and $f_{MB}$ are the splitting functions,
\begin{eqnarray}
f_{BM}(y)=\int_0^\infty|\Psi_{BM}(y,k_{\bot}^{2})|^{2}\d^{2} \emph{\textbf{k}}_{\bot}\;.
\end{eqnarray}
with the relation
\begin{eqnarray}
f_{MB}(y) = f_{BM}(1-y)\;.\label{sym}
\end{eqnarray}
The anti-quark numbers needed in this study can be obtained by integrating the anti-quark distribution given
by Eq.~(\ref{eq:qbarMCM}) over $x$.
The time-ordered perturbation theory~(TOPT) in the infinite
momentum frame~(IMF) is employed to calculate the splitting functions \cite{HolzenkampHS+Holtmann}
\begin{eqnarray}
f_{BM}(y)=\frac{1}{4\pi^{2}} \frac{m_{N}m_{B}}{y(1-y)}
\frac{|G_{M}(y,k_{\bot}^2)|^{2}|V_{\mathrm{IMF}}|^{2}}{[m_N^2-M_{BM}^2(y, k_{\bot}^2)]^2}\;,
\end{eqnarray}
where
\begin{eqnarray}
M_{BM}^2(y,k_{\bot}^2)=\frac{m_B^2+k_{\bot}^2}{y}+\frac{m_M^2+k_{\bot}^2}{1-y}
\end{eqnarray}
is the invariant mass squared of the final state, and $G_{M}(y,k_{\bot}^2)$ is a phenomenological
vertex form factor \cite{{HolzenkampHS+Holtmann,KoepfFS}},
\begin{eqnarray}
G_{M}(y,k_{\bot}^2)=\exp \left[ \frac{m_N^2-M_{BM}^2(y,k_{\bot}^2)}{2\Lambda_M^2} \right]\;.
\end{eqnarray}
A unique cutoff parameter $\Lambda_M=880$~MeV was taken in the calculation.

The Fock states considered include $| N \pi \rangle$, $ | N \rho \rangle$, $| N \omega \rangle$,
$ | \Delta \pi \rangle$, $ | \Delta \rho \rangle$,
$ | \Lambda K \rangle$, $ | \Lambda K^* \rangle$,
$ | \Sigma K \rangle$, and $ | \Sigma K^* \rangle$.
The vertex function $V_{IMF}^{\lambda\lambda^\prime}(y,k_\perp^2)$
depends on the effective interaction Lagrangian that describes the
fluctuation process $N \ra B M$. From the meson exchange model for
hadron production \cite{HolzenkampHS+Holtmann}, we have
\begin{eqnarray}
\emph{L}_1 &=& i g {\bar N}\gamma_5 \pi B\,,\nonumber \\
\emph{L}_2 &=&  f {\bar N} \partial_\mu \pi {\Delta}^\mu + \mbox{h.c.}\,, \nonumber \\
\emph{L}_3 &=&  g {\bar N} \gamma_\mu\theta^\mu B
                 +f {\bar N} \sigma_{\mu\nu}
        B (\partial^\mu\theta^\nu-\partial^\nu\theta^\mu)\,, \nonumber \\
\emph{L}_4 &=&i  f {\bar N} \gamma_5 \gamma_\mu \Delta_\nu
    (\partial^\mu\theta^\nu-\partial^\nu\theta^\mu) + \mbox{h.c.}\,,
\label{langragians}
\end{eqnarray}
where $N$ and $B=\Lambda,~\Sigma$ are spin-1/2 fields, $\Delta$ a
spin-3/2 field of Rarita-Schwinger form~($\Delta$ baryon), $\pi$ a
pseudoscalar field~($\pi$ and $K$), and $\theta$ a vector
field~($\rho$, $\omega$ and $K^*$). The coupling constants for
various fluctuations are taken to be \cite{HolzenkampHS+Holtmann,meson}
\begin{eqnarray}
g^2_{NN\pi}/4\pi = 13.6\,,& &  \nonumber \\
g^2_{NN\rho}/4\pi=0.84, & & f_{NN\rho}/g_{NN\rho}=6.1/4m_N\,,\nonumber \\
g^2_{NN\omega}/4\pi=8.1, & & f_{NN\omega}/g_{NN\omega}=0\,,\nonumber \\
f^2_{N\Delta\pi}/4\pi=12.3~{\rm GeV}^{-2}\,, & &
f^2_{N\Delta\rho}/4\pi=34.5~{\rm GeV}^{-2}\,, \nonumber \\
g_{N\Lambda K} =  -13.98\,, & &
g_{N\Sigma K}  =   2.69\,,  \nonumber \\
g_{N\Lambda K^{*}}  = -5.63\,, & &
f_{N\Lambda K^{*}}  = -4.89~{\rm GeV}^{-1}\,, \nonumber \\
g_{N\Sigma K^{*}}  = -3.25\,, & & f_{N\Sigma K^{*}}  = 2.09~{\rm
GeV}^{-1}\,. \label{couplingconstant}
\end{eqnarray}


\section{Numerical results and discussion}

Strange quark numbers from contributions of different mesons are presented
in Table~\ref{Table_strange}.
The first row is the CQM calculation under SU(3) symmetry assumption (i.e.  $\alpha=\beta=1$)
and with  $a=0.10$ and $\zeta=-1.2$ according to \cite{cl1}.
The second and third rows are the results when the SU(3) symmetry breaking effects are included using
the parameters of $\alpha$ and $\beta$ given in~\cite{song9705}.
The U(1) breaking parameter $\zeta$ was set in
order to give a overall description of the data as illustrated in~\cite{config05}.
In the next two rows, the results from the second method of introducing the symmetry break effects in
the CQM by calculating the fluctuation probability with different cutoff values are given (see Eq.~(\ref{eq:qbar2})).
The last row is the calculations from the MCM.
It can be found from Table~\ref{Table_strange} that for the CQM calculations
the ratio of the contributions from $K$ and $\eta+\eta'$ depends strongly on the
parameters measuring the SU(3) breaking effects.
For the second method introducing symmetry breaking effects in the CQM calculation,
the strange quark number is very sensitive to the  cutoff parameter $\Lambda_\delta$.
The contribution from the $K$ meson calculated in the MCM is $3 \sim 10$ times smaller than that in
the CQM with symmetry breaking effects. Though the contribution from the $K^*$ meson
compensates to some extent, one can conclude that this two
models give very different results for the magnitude of the non-perturbative strange sea.
The  $\eta$ and $\eta'$ mesons play an
important role in the CQM calculation, while their contributions are negligible in the MCM calculation as
their coupling constants are usually taken to be zero in this model.

\begin{table}
\caption{Strange quark number from different mesons}
\label{Table_strange}
\begin{tabular}[t]{|c|c|c|c|c|}
\hline  $\bar{s}$& $K$& $\eta+\eta'$ & $K^*$ & sum
\\ \hline  CQM($\alpha=\beta=1.0$, $\zeta=-1.2$) & 0.30 & 0.16& -&0.46
\\ \hline  CQM($\alpha=0.4$, $\beta=0.7$, $\zeta=-0.65$) & 0.048 &
0.061&-&0.11
\\ \hline CQM($\alpha=\beta=0.45$, $\zeta=-0.1$) & 0.061 & 0.031  &
-&0.092
\\ \hline CQM($\Lambda_\delta=1500$ MeV) & 0.17 & 0.030
&-&0.20
\\ \hline CQM($\Lambda_\delta=1100$ MeV) & 0.092 & 0.016
&-&0.11
\\ \hline  MCM & 0.016 & - &  0.026&0.042
\\\hline
\end{tabular}
\end{table}

The results for $\sigma_{\pi N}$ are presented in Table~\ref{Table_yN} with indexes
1 and 2 referring to the calculations adopting $\hat{\sigma}=26$~MeV and $36$~MeV respectively.
For the CQM calculations with SU(3) symmetry being held,
the larger value of $\hat{\sigma}$ yields $\sigma_{\pi N}=67.8$~MeV which is
comparable with recent analyses of $55 \sim 75$ MeV \cite{Kaufmann:dd,Olsson:pi}.
If the SU(3) symmetry breaking effects are included, the strange quark number becomes much smaller and
$\sigma_{\pi N}$ decreases by $24 \sim 28$ MeV depending on the method of introducing
the SU(3) symmetry breaking effects,
which gives a value for  $\sigma_{\pi N}$ close to the earlier analyses of $45$~MeV \cite{Koch:pu,Gasser:1990ce}.
With the smaller value of $\hat{\sigma}$ the results for $\sigma_{\pi N}$ with the symmetry
breaking effects are smaller than the earlier analyses of $45$~MeV \cite{Koch:pu,Gasser:1990ce}
by about $15$ MeV.
The two methods of introducing the breaking effects in the CQM are consistent with each
other, and almost have the same impact on the $\sigma_{\pi N}$.
For the calculations in the MCM, the fluctuation
probability of $K^*$ is at the same level as $K$ \cite{Cao}. Consequently adding vector
meson $K^*$ will double the strange quark number, which changes the calculations of $f_s$ and
$y_N$ dramatically. However, the value obtained for the $\sigma_{\pi N}$ remains quite small.
Comparing the two model results, one can find that the strange quark number from the CQM
with the symmetry breaking effects is
$2 \sim 5$ times larger than that form the MCM, but the difference of
$\sigma_{\pi N}$ from the two models is relatively small (less than $20$ per cents).

\begin{table}
\caption{$y_N$ and $\sigma_{\pi N}$ with non-perturbative sea}
\label{Table_yN}
\begin{tabular}[t]{|c|c|c|c|c|c|}
 \hline  & $\bar{s}$& $f_s$ & $y_N$ & $\sigma_{\pi N}^1$ &$\sigma_{\pi N}^2$
\\ \hline\hline CQM($\alpha=\beta=1.0$, $\zeta=-1.2$) & 0.46 & 0.19 & 0.23 & 48.9 & 67.8
\\ \hline CQM($\alpha=0.4$, $\beta=0.7$, $\zeta=-0.65$) & 0.11 & 0.053 & 0.056 & 29.3 &
40.5
\\ \hline CQM($\alpha=\beta=0.45$, $\zeta=-0.1$) & 0.092 & 0.042 & 0.044 & 28.5 &
39.5
\\ \hline CQM($\Lambda=1500$ MeV) & 0.20 & 0.083 & 0.09 & 31.7 & 44.0
\\ \hline CQM($\Lambda=1100$ MeV) & 0.11 & 0.047 & 0.050 & 28.9 & 40.0
\\ \hline\hline MCM(only K) & 0.016 & 0.0075 & 0.0075 & 26.3 & 36.6
\\ \hline MCM(add $K^*$) & 0.042 & 0.020 & 0.02 & 27.1 & 37.5
\\\hline
\end{tabular}
\end{table}

\begin{table}
\caption{$\sigma_{K N}$ and $\sigma_{\eta N}$ with non-perturbative sea.}
\label{Table_KN}
\begin{tabular}[t]{|c|c|c|c|c|c|c|c|}
 \hline  & $y_N$ & $\sigma_{\pi N}^1$ & $\sigma_{\pi N}^2$ & $\sigma_{K N}^1$
& $\sigma_{K N}^2$& $\sigma_{\eta N}^1$ & $\sigma_{\eta N}^2$
\\ \hline\hline CQM($\alpha=\beta=1.0$, $\zeta=-1.2$)& 0.23 & 48.9 & 67.8 & 464.1
&643.4&391.2&542.4
\\ \hline CQM($\alpha=0.4$, $\beta=0.7$, $\zeta=-0.65$) & 0.056 &
29.3 & 40.5&211.8&292.7&64.5&89.1
\\ \hline CQM($\alpha=\beta=0.45$, $\zeta=-0.1$) & 0.044 & 28.5
&39.5&201.6& 279.3&51.3&71.1
\\ \hline CQM($\Lambda=1500$ MeV) & 0.09 & 31.7 &44.0&243.1&337.5&105.7&146.7
\\ \hline CQM($\Lambda=1100$ MeV) & 0.050 & 28.9 &40.0&206.6&286.0&57.8&80.0
\\ \hline\hline MCM(only K) & 0.0075 & 26.3 &36.6&173.5&241.5&15.3&21.4
\\ \hline MCM(add $K^*$) & 0.02 & 27.1 &37.5&183.2&253.5&27.1&37.5
\\\hline
\end{tabular}
\end{table}

The results for $\sigma_{K N}$ and $\sigma_{\eta N}$ are presented in Table \ref{Table_KN}.
The calculations for the $\sigma_{K N}$ and $\sigma_{\eta N}$ are more sensitive
to $y_N$ than that for the $\sigma_{\pi N}$.
From Table~\ref{Table_KN} one can find that the CQM calculations for the $\sigma_{K N}$
and $\sigma_{\eta N}$ with the SU(3) breaking effects are about 2 and 4 $\sim$ 8 times smaller than
that when the effects are not included.
The MCM calculations for $\sigma_{K N}$ and $\sigma_{\eta N}$ are about $10\%$ and
$50\%$ smaller that the smallest results from the CQM while two models, as discussed above,
may give comparable results for the $\sigma_{\pi N}$.

There are much less calculations for the $\sigma_{K N}$ and $\sigma_{\eta N}$ than
for the $\sigma_{\pi N}$. A theoretical study based on the lattice QCD calculation
gave $\sigma_{K N}=362 \pm 13$~MeV~\cite{Dong} and
an analysis using the Nambu-Jona-Lasinio model predicated
$\sigma_{KN}=425$~MeV (with an error of $10 \sim 15\%$) \cite{Hatsuda_Kunihiro}.
A calculation using the perturbative chiral quark model gave $ \sigma_{KN} =
312 \pm 37 \, {\rm MeV} $ and $\sigma_{\eta N}=72 \pm 16$~MeV \cite{SigmaTerm01}.
Our calculations using the CQM with symmetry breaking effects are comparable with these results,
while the results from the MCM are considerable smaller than these results.
Future DA$\Phi$NE experiments~\cite{Gensini} will allow for a determination of the
$K N$ sigma terms and hence give a more narrow range of the nucleon strangeness.

The antiquark numbers given by Eqs.~(\ref{eq:qbar1}) and (\ref{eq:qbar2}) in the CQM
and the antiquark distribution given by Eq.~(\ref{eq:qbarMCM}) in the MCM come from non-perturbative effects.
The contributions from the process of gluons perturbatively splitting into quark-antiquark
pairs need to be treated carefully, although any effect associated with gluon is expected to be
small in the CQM.
In several recent studies of $x$ dependence of the sea quark distributions in the MCM, this effect was included
by using a phenomenal  parameterization for the symmetric nucleon sea in the bare
nucleon \cite{HN,gluon}.
However, the perturbative quark distributions are divergent as $x \ra 0$, thus the quark
and antiqaurk numbers from gluon perturbatively splitting can not be estimated using the same method.
Further investigations  are highly needed in order to include these perturbative contributions.

\section{Summary}

We calculated the nucleon strangeness $y_N$  in the chiral quark model and the meson cloud model,
and used $y_N$ as a parameter to evaluate the the nucleon sigma terms ($\sigma_{\pi N}$, $\sigma_{K N}$
and $\sigma_{\eta N}$).
Our calculations show that $y_N$ from the CQM is much larger than that from the MCM,
while the difference for $\sigma_{\pi N}$ between the two models is
relatively small. The results indicate that adopting a larger value of $\hat{\sigma}$
from higher order calculations in ChPT can give a value of $\sigma_{\pi N}$ comparable with the earlier analyses.
The only value that could be comparable to the recent analyses using the $\pi N$ scattering data
that gave a value in the range of $55 \sim 75$ MeV
is the result from the CQM with SU(3) symmetry.
This picture, however, was not a good description of light flavor
antiquark asymmetry and the quark spin component. Meanwhile, the
results of $\sigma_{K N}$ and $\sigma_{\eta N}$ become strangely
large with symmetry being held, which confirms that the SU(3) symmetry
breaking effects should be considered.
The higher value of $\sigma_{\pi N}$ from recent analyses of pion-nucleon scattering data
needs to be re-considered carefully. Another important conclusion is
that both $\sigma_{K N}$ and $\sigma_{\eta N}$ are quite sensitive
to $y_N$. Thus, more exact values of them determined from the future
experiments could restrict the model parameters and provide a better
knowledge of the strangeness content of the nucleon.

\section*{Acknowledgments}

This work is partially supported by the Marsden Fund of the Royal
Society of New Zealand, by National Natural Science Foundation of
China~(Nos.~10421503, 10575003, 10528510), by the Key Grant
Project of Chinese Ministry of Education~(No.~305001), and by the
Research Fund for the Doctoral Program of Higher
Education~(China).


\begin{thebibliography}{99}

\bibitem{ST} A. I. Signal and A. W. Thomas, Phys. Lett. B {\bf 191}, 205 (1987).
\bibitem{bm96} S. J. Brodsky and B.-Q. Ma, Phys. Lett. B {\bf 381}, 317 (1996).
\bibitem{zell02} G. P. Zeller $et~al.$, Phys. Rev. Lett. {\bf 88}, 091802 (2002);
                 Phys. Rev. D {\bf 65}, 111103(R) (2002).
\bibitem{dm04} Y. Ding and B.-Q. Ma, Phys. Lett. B {\bf 590}, 216 (2004).
\bibitem{ding}
    Y. Ding, R.-G. Xu, and B.-Q. Ma, Phys. Lett. B {\bf 607}, 101 (2005);
    Y. Ding, R.-G. Xu, and B.-Q. Ma, Phys. Rev. D {\bf 71}, 094014 (2005).
\bibitem{wakamastu04} M. Wakamatsu, Phys. Rev. D {\bf 71}, 057504 (2005).
\bibitem{Cao}F.-G. Cao and A. I. Signal, Phys. Lett. B \textbf{559}, 229 (2003).
\bibitem{TPCheng1976} T. P. Cheng, \PRD {\bf 13}, 2161 (1976).
\bibitem{review-early} E.~Reya, Rev.\ Mod.\ Phys.\  {\bf 46}, 545 (1974); R.~L.~Jaffe, Phys.\ Rev.\ D {\bf 21} 3215 (1980).
\bibitem{review-recent}J.~Gasser and M.~E.~Sainio,
in {\sl Proceedings of the 3rd Workshop on Physics and Detectors for DAPHNE (DAPHNE 99), Frascati, Nov. 1999},
edited by S. Bianco, F. Bossi, G. Capon, F. L. Fabbri, P. Gianotti, G. Isidori, and F. Murtas (Istituto Naz. Fis. Nucl., 1999), p. 659;
 M.~E.~Sainio, $\pi$N Newsletter {\bf 16}, 138 (2002).
\bibitem{g1}
        J.~Gasser, Ann. Phys.~(N.Y.) {\bf 136}, 62 (1981); J.~Gasser and H.~Leutwyler, Phys. Rep. {\bf 87}, 77 (1982).
\bibitem{bm}
       B.~Borasoy and U.-G.~Mei{$\beta$}ner, Ann. Phys.~(N.Y.) {\bf 254}, 192 (1997).
\bibitem{Koch:pu}
  R.~Koch, Z.\ Phys.\ C {\bf 15}, 161 (1982).


\bibitem{Gasser:1990ce}
  J.~Gasser, H.~Leutwyler, and M.~E.~Sainio,
  Phys.\ Lett.\ B {\bf 253}, 252;  {\bf 253}, 260 (1991).
\bibitem{Kaufmann:dd}
  W.~B.~Kaufmann and G.~E.~Hite,
  Phys.\ Rev.\ C {\bf 60}, 055204 (1999); M.~G.~Olsson,
  Phys.\ Lett.\ B {\bf 482}, 50 (2000); M.~M.~Pavan, I.~I.~Strakovsky, R.~L.~Workman, and R.~A.~Arndt,
  $\pi$N Newsletter {\bf 16}, 110 (2002).
\bibitem{Olsson:pi}
  M.~G.~Olsson and W.~B.~Kaufmann,
  $pi N$ Newsletter  {\bf 16}, 382 (2002).
\bibitem{Leinweber:2003dg}
   D.~B.~Leinweber, A.~W.~Thomas, and R.~D.~Young, \PRL {\bf 92}, 242002 (2004).
\bibitem{MProcuraHW}
    M. Procura, T. R. Hemmert, and W. Weise, \PRD {\bf 69}, 034505 (2004).

\bibitem{SigmaTerm01}
    T. Inoue, V. E. ~Lyubovitskij, Th.~Gutsche, and A.~Faessler, \PRC {\bf 69}, 035207 (2004);
    V.~E.~Lyubovitskij, Th.~Gutsche, A.~Faessler, and E.G. Drukarev, \PRD {\bf 63}, 054026 (2001).

\bibitem{ratio25}
    J.~Gasser, Ann. Phys. {\bf 136}, 62 (1981); H.~Leutwyler, Phys. Lett. B {\bf 378}, 313 (1996).
\bibitem{cl1}
        T.~P.~Cheng and L.-F.~Li, \PRL {\bf 74}, 2872 (1995).
\bibitem{cl2}
        T.~P.~Cheng and L.-F.~Li, Phys. Rev. D {\bf 57}, 344 (1998).
\bibitem{mg}
        A.~Manohar and H.~Georgi, \NPB {\bf234}, 189 (1984).
\bibitem{ehq92}
        E.~J.~Eichten, I.~Hinchliffe, and C.~Quigg, \PRD {\bf 45}, {2269} (1992).
\bibitem{AJBuchmannF96}
A. Szczurek, A. J. Buchmann, and A. Faessler, J. Phys. G {\bf 22},
1741 (1996).
\bibitem{smw}
    X.~Song, J.~S.~McCarthy, and H.~J.~Weber, \PRD {\bf 55}, 2624 (1997).
\bibitem{wsk}
    H.~J.~Weber, X.~Song, and M. Kirchbach, Mod. Phys. Lett. A {\bf 12}, 729 (1997).
\bibitem{song9705}
    X.~Song, Phys. Rev. D {\bf 57}, 4114 (1998).
\bibitem{sw98}
   K. Suzuki and W. Weise, Nucl. Phys. A {\bf634}, 141 (1998).
\bibitem{w90} S. Weinberg, Phys. Rev. Lett. {\bf65}, 1181 (1990).
\bibitem {first} A. W. Thomas, Phys. Lett. B {\bf 126}, 97 (1983).
\bibitem {first2} E. M. Henley and G. A. Miller, Phys. Lett. B {\bf 251}, 453 (1990).
\bibitem {MA} M. Alberg, T. Falter, and E.M. Henley, Nucl. Phys. A {\bf 644}, 93 (1998).
\bibitem {Review}
For recent reviews, see, {\it e.g.}, S. Kumano, Phys. Rep. {\bf 303}, 183 (1998);
J. Speth and A.~W.~Thomas, Adv. Nucl. Phys {\bf 24}, 83 (1998);
G. T. Garvey and J.C. Peng, Prog. Part. Nucl. Phys. {\bf 47}, 203 (2001).
\bibitem{KoepfFS} W.~Koepf, L.L.~Frankfurt, and M.~Strikman, Phys. Rev. D {\bf 53}, 2586 (1996).
\bibitem{HolzenkampHS+Holtmann}
B. Holzenkamp, K. Holinde and J. Speth, \Journal{\NPA}{500}{485}{1989};
H.~Holtmann, A.~Szczurek, and J. Speth, \Journal{\NPA} {569}{631}{1996}.
\bibitem {meson} R. Machleidt, K. Holinde, and Ch. Elster, Phys. Rep. {\bf 149}, 1 (1987).

\bibitem{config05} H.~Dahiya, M.~Gupta, and J.M.S.~Rana,
Int. J. Mod. Phys. A {\bf 21}, 4255 (2006)
\bibitem{Dong}S.~J.~Dong, J.-F.~Laga\"{e}, and K.~F.~Liu,
Phys. Rev. D {\bf 54}, 5496 (1996).
\bibitem{Hatsuda_Kunihiro} T.~Hatsuda and T.~Kunihiro,
Phys. Rep. {\bf 247}, 221 (1994).
\bibitem{Gensini} P.~M.~Gensini,
R.~Hurtado, and G.~Violini, arXiv:hep-ph/9804344.

\bibitem {HN} M. Alberg and E. M. Henley, Nucl. Phys. {\bf A663}, 301 (2000).
\bibitem {gluon} A. Vogt,
in  {\sl Proceedings of the 6th International Workshop On Deep Inelastic Scattering and QCD (DIS 98),  Brussels, April 1998},
edited by Gh. Coremans and R. Roosen (World Scientific,  Singapore, 1998), p. 792.
\end{thebibliography}
\end{document}